\documentclass[11pt]{article}
\usepackage{amssymb}
\def\ba{\mathbf{a}}   
\def\bb{\mathbf{b}}   
\def\bc{\mathbf{c}}   
\def\be{\mathbf{e}}   
\def\bm{\mathbf{m}}   
\def\bn{\mathbf{n}}   
\def\bv{\mathbf{v}}   
\def\bx{\mathbf{x}}   

\def\bA{\mathbf{A}}   
\def\bM{\mathbf{M}}

\def\C{\mathbb{C}} 
\def\G{\mathbb{G}} 

\def\R{\mathbb{R}}  

\def\no{\noindent}

\def\beq{\begin{equation}}
\def\eeq{\end{equation}}

\def\w{\wedge}

\usepackage{helvet}
\usepackage{courier}
\usepackage{graphicx}   
\usepackage{makeidx}
\usepackage{multicol} 
\usepackage{mathptmx} 
\def\bpm{\begin{pmatrix}}
\def\epm{\end{pmatrix}}
\usepackage[margin=2cm]{geometry}
\makeindex            
\begin{document}
\title{Geometric Number Systems and Spinors}
\author{Garret Sobczyk
\\ Universidad de las Am\'ericas-Puebla
 \\ Departamento de F\'isico-Matem\'aticas
\\72820 Puebla, Pue., M\'exico
\\ http://www.garretstar.com}
\maketitle
\begin{abstract} The real number system is geometrically extended to include three new {\it anticommuting} square roots of plus one, each such root representing the direction of a unit vector along the
orthonormal coordinate axes of Euclidean 3-space. 
The resulting geometric (Clifford) algebra provides a geometric basis for the famous Pauli matrices which,
in turn, proves the consistency of the rules of geometric algebra. 
 The flexibility of the concept of geometric numbers opens the
door to new understanding of the nature of space-time, and of Pauli and Dirac spinors 
as points on the Riemann sphere, including Lorentz boosts.

\smallskip
\no {\em AMS Subject Classification:} 15A66, 81P16
\smallskip

\no {\em Keywords:} geometric algebra, relative geometric algebra,  
 Riemann sphere, complex Riemann sphere.

\end{abstract}

\section{The geometric algebra of space $\G_3$}

\begin{figure}
\begin{center}
\includegraphics[scale=.25]{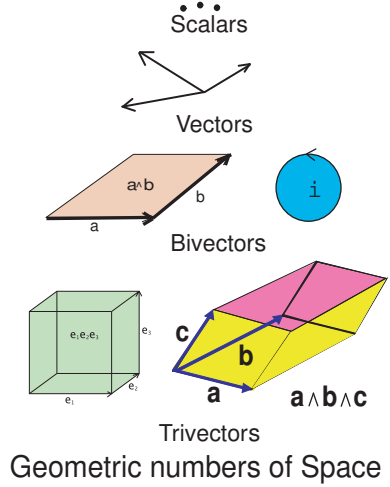}
\caption{Geometric numbers of space}
\label{picbasis}
\end{center}
\end{figure}

The most direct way of obtaining the geometric algebra $\G_3$ of space is to extend the
real number system $\R$ to include three new {\it anti-commutative} square roots
$\be_1, \be_2 , \be_3$ of $+1$, that represent {\it unit vectors} along the respective $xyz$-coordinate axes. 
Thus, $\be_1,\be_2, \be_3 \not \in \R$ and $\be_1^2=\be_2^2=\be_3^2=1$.
The resulting associative geometric algebra $\G_3:=\R[\be_1,\be_2,\be_3]$, as a real linear space, has
the $2^3=8$-dimensional {\it standard basis}
\[  \G_3 = span_\R\{ 1, \be_1,\be_2,\be_3, \be_{12},\be_{13},\be_{23},\be_{123}\}, \]
where $\be_{jk}:=\be_j \be_k = -\be_k \be_j$ represent unit {\it bivectors} in the three 
$xy, xz, yz$-coordinate planes, for $j\ne k$, and $I:=\be_{123}=\be_1\be_2 \be_3$ represents the oriented {\it directed  trivector}, or {\it pseudoscalar} element of space. 
The geometric numbers of $3$-dimensional space are pictured in Figure \ref{picbasis}.

To see that the rules of our {\it geometric algebra} are consistent, we relate it immediately to the famous
Pauli algebra ${\cal P}$ of square $2\times 2$ matrices over the complex numbers $\C$. The most intuitive
way of doing this is to introduce the {\it mutually annihilating idempotents} $u_\pm = \frac{1}{2}(1 \pm \be_3)$,
which satisfy the rules
\[  u_+^2 = u_+, \ \ u_-^2=u_-, \ \ u_+ u_- = 0, \ \ u_++u_- = 1, \ \ u_+-u_- = \be_3. \]
In addition, $\be_1 u_+ = u_- \be_1$.  All these rules are easily verified and left to the reader. Another important
property of the geometric algebra $\G_3$ is that the pseudoscalar element $I=\be_{123}$ is in the {\it center} of the 
algebra, commuting with all elements, and $I^2=-1$. Thus $I$ can take over the roll of the unit imaginary
$i=\sqrt{-1}$.

By the {\it spectral basis} $ {\cal S}_\G$  of the geometric algebra $\G_3$, we mean
\beq {\cal S}_\G:= \pmatrix{1 \cr \be_1}u_+ \pmatrix{1 & \be_1} =\pmatrix{ u_+ & \be_1 u_- \cr
                                   \be_1 u_+ & u_-} ,\label{spectralbasis} \eeq
where $I$ is taking over the roll of $i=\sqrt{-1}$ in the usual Pauli algebra.
Any geometric number $g = s+It + \ba +I  \bb \in \G_3$ corresponds directly to a Pauli matrix $[g]$,
by the simple rule
\[  g =  \pmatrix{1 & \be_1}u_+[g] \pmatrix{1 \cr \be_1} .\]
For example, the famous Pauli matrices $[\be_1], [\be_2], [\be_3]$ are specified by
\beq  [\be_1]:= \pmatrix{0 & -1 \cr 1 & 0},  \ \  [\be_2]:= \pmatrix{0 & -I \cr I & 0}, \ \ 
 [\be_1]:= \pmatrix{1 & 0 \cr 0 & -1},   \label{paulimatrices} \eeq
as can be easily checked. For $[\be_2]$, we have
  \[  \be_2 = \pmatrix{1 & \be_1}u_+ \pmatrix{0 & -I \cr I & 0}\pmatrix{1 \cr \be_1} = I \be_1 (u_+-u_-) =I \be_1 \be_3. \]

This shows that the Pauli algebra ${\cal P}$ and the geometric algebra $\G_3$ are fully compatible as
algebraically isomorphic algebras over the complex numbers 
\[ \C:=\{z=s+t I| \ \ {\rm for} \ \  s,t \in \R\}.  \]
 The geometric product of geometric numbers $g_1g_2$ corresponds to the usual matrix algebra product
of the matrices $[g_1][g_2]=[g_1g_2]$. The great advantage of the geometric algebra $\G_3$ over the
Pauli algebra $\cal P$, is that the geometric numbers are liberated from their coordinate representations as
$2\times 2$ complex matrices, as well as being endowed with a complete geometric interpretation. On the
other hand, the consistency of the rules of the geometric algebra follow from the known consistency of rules
of matrix algebra, and matrices offer a computational tool for computing the product of geometric numbers
\cite{S08}.

It is worthwhile to give a summary of the deep relationship between pre-relativistic (Gibbs-Heaviside) vector algebra,
and the geometric algebra $\G_3$. The geometric product of two vectors $\ba, \bb \in \G_3^1$ is given by
\beq  \ba  \bb = \frac{1}{2}(\ba \bb + \bb \ba)  + \frac{1}{2}(\ba \bb - \bb \ba)  = \ba \cdot \bb + \ba \w \bb ,\label{geoprodvecs} \eeq
where the {\it inner product} $\ba \cdot \bb :=  \frac{1}{2}(\ba \bb + \bb \ba) \in \R $, and the {\it outer product}
$\ba \w \bb :=  \frac{1}{2}(\ba \bb - \bb \ba)  $ has the interpretation (due to Grassmann) of the {\it bivector}
in the plane of the vectors $\ba$ and $\bb$. The bivector $\ba \w \bb = I (\ba \times \bb)$, where
$\ba \times \bb$ is the vector normal to the plane of $\ba \w \bb$, and is its {\it dual}.

Another advantage of the geometric product $\ba \bb$ over the inner product $\ba\cdot \bb$ and the 
cross product $\ba \times \bb$, is the powerful cancellation rule
\[ \ba \bb = \ba \bc \quad \iff \quad \ba^2 \bb = \ba^2 \bc \quad \iff  \bb = \bc, \]
provided of course that $\ba^2 \ne 0$. It takes knowledge of {\it both}  $\ba\cdot \bb$ and $\ba\w \bb$  (or $\ba \times \bb$),
to uniquely determine the relative directions of the vectors $\ba$ and $\bb$. Another unique advantage
in the the geometric algebra is the {\it Euler formula} made possible by (\ref{geoprodvecs}), 
\[  \ba \bb =|\ba |\bb|e^{I\hat \bc \theta} =|\ba||\bb|(\cos \theta +I\hat \bc \sin \theta),\]
where $|\ba|:=\sqrt{\ba^2}$ and similarly for $|\bb|$, and $\theta$ is the angle between the
vectors $\ba$ and $\bb$. The unit vector $\hat \bc$ can be defined by $\hat \bc = \frac{\ba \times \bb}{|\ba \times \bb|}$.

The Euler formula for the bivector $I \hat \bc$, which has square $-1$, is the generator of rotations in the plane
of the bivector $\ba \w \bb$. Later, when talking about {\it Lorentz boosts}, we will also utilize the {\it hyperbolic
Euler form}
\beq e^{\phi \hat \bv}= \cosh \phi + \hat \bv \sinh \phi, \label{hypeulerform} \eeq
where $\tanh \phi =\frac{v}{c}\in \R$ determines the {\it rapidity} of the boost in the direction of the unit vector
$\hat \bv \in \G_3^1$.

A couple more formulas, relating the vector cross and dot products to the geometric product, are
  \[  \ba \w \bb \w \bc := \ba \cdot (\bb \times \bc) I = I\det[{\ba,\bb,\bc}],  \] 
and
  \[ \ba \cdot (\bb \w \bc):=(\ba \cdot \bb)\bc-(\ba \cdot \bc) \bb = -\ba \times (\bb \times \bc). \]
The triple vector products $\ba \w \bb \w \bc$ and $ \ba \cdot (\bb \w \bc)$ are directly related to the
geometric product by the identity
\[ \ba (\bb \w \bc) = \frac{1}{2}\Big(\ba(\bb\w \bc)-(\bb \w \bc)\ba\Big) +\frac{1}{2}\Big(\ba(\bb\w \bc)+(\bb \w \bc)\ba\Big), \]
where $ \ba \cdot (\bb \w \bc)= \frac{1}{2}\Big(\ba(\bb\w \bc)-(\bb \w \bc)\ba\Big)$ and
$ \ba \w (\bb \w \bc)= \frac{1}{2}\Big(\ba(\bb\w \bc)+(\bb \w \bc)\ba\Big)$. 
Detailed discussions and proofs of these identities, and their generalizations to higher dimensional geometric algebras, can be found in \cite{SNF}, \cite{S1}. Geometric algebra has in recent years become a basic tool for research in quantum mechanics
 \cite{BK12}, \cite{H10}, and more generally as a basic language of mathematics and physics \cite{H99}, \cite{SNF}.

We now return to beautiful results which depend in large part only upon the geometric product. Thus the reader can
relax and rely upon the familiar rules of matrix algebra, which are equally valid in the isomorphic geometric algebra.

\section{Stereographic projection in $\R^3$}

Consider the equation 
\beq \bm = \frac{2}{\hat \ba + \be_3}= \frac{2(\hat \ba + \be_3)}{(\hat \ba + \be_3)^2}
= \frac{\hat \ba + \be_3}{1+\hat \ba \cdot \be_3} , \label{ahateqn} \eeq
where $\hat \ba =\ba/|\ba|$ is a unit vector for the vector $\ba \in \G_3^1$. Clearly,
this equation is well defined except when $\hat \ba = - \be_3$. Let us solve this equation
for $\hat \ba$, but first we find that
\[ \bm\cdot \be_3 =\Big( \frac{2}{\hat \ba + \be_3}\Big)\cdot \be_3=1. \]
Returning to equation (\ref{ahateqn}),
\beq  \hat \ba = \frac{2}{\bm}-\be_3= \frac{1}{\bm}\big(2- \bm \be_3 \big)
                       = \frac{1}{\bm}\big(2+\be_3 \bm -2 \be_3\cdot \bm \big)
=\hat \bm \be_3 \hat \bm   . \label{ahateqn2} \eeq
Equation (\ref{ahateqn}) can be equivalently expressed by
\[  \hat \ba=\hat \bm \be_3 \hat \bm =(\hat \bm \be_3)\be_3(\be_3 \hat \bm) =(-I\hat \bm) \be_3 (I\hat \bm) , \]
showing that $\hat \ba$ is obtained by a rotation of $\be_3$ in the plane of $\hat \bm\w \be_3$ through an
angle of $2\theta$ where $\cos\theta:=\be_3\cdot \hat \bm$, or equivalently,
by a rotation of $\be_3$ in the plane of $I \hat \bm$ through an angle of $\pi$	.

It is easily shown that the most general idempotent in $\G_3$ has the form
\beq  s= \frac{1}{2}(1+\bm + I \bn) \label{mnidempotent} \eeq
where
\[ (\bm+I \bn)^2 = 1 \quad \iff \quad \bm^2- \bn^2 = 1 \ \ {\rm and} \ \ \bm \cdot \bn = 0, \]
and $I:= \be_{123}$ is the unit pseudoscalar element in $\G_3^3$.
Consider now idempotents of the form $p=(1+ \lambda \be_1)u_+$, where $\lambda \in \G_3^{0+3}$.
Equating $s = p$, we find that
\[  (1+\bm + I \bn) =(1+ \lambda \be_1)(1+ \be_3) =1+\lambda \be_1 - I\lambda \be_2 + \be_3 .   \]
Changing the parity of this equation, gives
\[  (1-\bm + I \bn) =(1- \lambda^\dagger \be_1)(1- \be_3) =1-\lambda^\dagger \be_1 - I \lambda^\dagger \be_2 - \be_3 ,   \]
since the {\it parity change} (changing the sign of vectors) $\lambda^-$ of $\lambda$ is identical to the {\it reverse} (reversing the order of products of vectors) $\lambda^\dagger$ of $\lambda$.

We can now solve these last two equations for $\bm$ and $I\bn$ in terms of $\lambda$, getting 
\[ \bm = \frac{\lambda+\lambda^\dagger}{2}\be_1+\frac{\lambda-\lambda^\dagger}{2I}\be_2+\be_3, \quad
           I \bn =\frac{\lambda-\lambda^\dagger}{2}\be_1-\frac{\lambda+\lambda^\dagger}{2I} \be_2 = \bm \w \be_3,\]
or
\beq \bm = \bx+\be_3 \quad {\rm and} \quad \bn= \bm \times \be_3, \label{mnbx}  \eeq
where $\bx=x\be_1+ y\be_2 \in \R^2$, the $xy$-plane.
From (\ref{mnidempotent}), it immediately follows that
\beq s = \frac{1}{2}(1+\bm+I\bn)=\frac{1}{2}(1+\bm+\bm\w \be_3)=\bm u_+ .\label{canonicals} \eeq
We also easily find that
\[ \bm^2= 1+\lambda \lambda^\dagger =1+\bx^2  \ge 1 \quad \leftrightarrow \quad |\bm| = \sqrt{1+\lambda \lambda^\dagger}=\sqrt{1+\bx^2}. \]

A {\it Pauli spinor} is a column matrix of two complex components, which we denote by
 $[\alpha]_2:=\pmatrix{\alpha_0 \cr \alpha_1}$.
Each Pauli spinor $[\alpha]_2$ corresponds to a {\it minimal left ideal} $[\alpha]_L$, which in turn corresponds to  {\it geometric Pauli spinor}, or {\it Pauli g-spinor} $\alpha$ in the geometric algebra $\G_3$. We have
\beq  [\alpha]_2=\pmatrix{\alpha_0 \cr \alpha_1} \quad \longleftrightarrow \quad [\alpha ]_L := \pmatrix{\alpha_0 & 0 \cr \alpha_1 & 0} 
\quad \longleftrightarrow \quad \alpha := (\alpha_0 +\alpha_1 \be_1)u_+\in \G_3.\label{spinorcorespondence} \eeq
By factoring out $\alpha_0$ from $g$-spinor $\alpha$, we get
\[  \alpha =  (\alpha_0 +\alpha_1 \be_1)u_+ = \alpha_0 (1+ \frac{\alpha_1}{\alpha_0}\be_1)u_+= \alpha_0 p = \alpha_0 \bm u_+,\]
where $p=s$ is the idempotent defined above for $\lambda = \frac{\alpha_1}{\alpha_0}$ .

By the {\it norm} $|\alpha|$ of the $g$-spinor $\alpha$, we mean
\beq |\alpha|:=\sqrt{2\langle \alpha^\dagger \alpha\rangle_0}= \sqrt{\alpha_0\alpha_0^\dagger+\alpha_1\alpha_1^\dagger} \ge 0 ,                  \label{spinornorm} \eeq
where $\langle g \rangle_0$ means the {\it real number part} of the geometric number $g\in \G_3$. More generally
we define the {\it sesquilinear inner product} between the $g$-spinors $\alpha, \beta$ to be
\[  \langle \alpha | \beta \rangle := 2\langle \alpha^\dagger \beta \rangle_{0+3} = \alpha_0^\dagger \beta_0+\alpha_1^\dagger \beta_1, \]
where $\langle g \rangle_{0+3}$ means the {\it scalar and pseudo-scalar parts} of the geometric number $g\in \G_3$. A $g$-spinor $\alpha$ is said to be {\it normalized} if $|\alpha|=1$.

\begin{figure}
\begin{center}
\no\includegraphics[scale=.30]{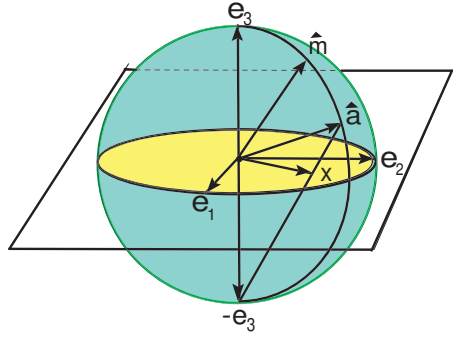}
\caption{Stereographic Projection from the South Pole to the $xy$-plane. }
\label{sterox}
\end{center}
\end{figure}

From equations (\ref{mnidempotent}) and (\ref{spinornorm}), it follows that for $\lambda=\alpha_0^{-1}\alpha_1$,
\[ \alpha = \alpha_0 s = \alpha_0 \sqrt{1+\lambda \lambda^\dagger}\hat \bm u_+=\rho e^{I\theta} \hat \bm u_+
= \rho e^{I\theta}\hat \ba_+ \hat \bm, \]
where $e^{I \theta}:=\frac{ \alpha_0}{ \sqrt{\alpha_0\alpha_0^\dagger}}$, $\rho :=\sqrt{\alpha_0\alpha_0^\dagger
+\alpha_1 \alpha_1^\dagger}$, and $\hat \ba_+:= \hat \bm u_+ \hat \bm$. 
Equations (\ref{ahateqn}) and (\ref{ahateqn2})
have an immediate interpretation on the Riemann sphere centered at the origin.  Figure \ref{sterox} shows a cross-section of the Riemann $2$-sphere,
taken in the plane of the bivector $\bm \w \be_3$, through the origin. We see that the stereographic projection 
from the South pole at the point $-\be_3$, to the point $\hat \ba$ on the Riemann sphere,
passes through the point $\bx=proj(\bm)$ of the point $\bm$ onto the plane through the origin with the normal vector $\be_3$.
Stereographic projection is
 just one example of conformal mappings, which have important generalizations to higher dimensions \cite{Sob2012}. 
  
We can now simply answer a basic question in quantum mechanics. If a spin $\frac{1}{2}$-particle is prepared in
a normalized Pauli $g$-spin state $\alpha$, what is the probability of finding it in a normalized Pauli g-state $\beta$ immediately
thereafter?
We calculate
\[  \langle \beta |  \alpha \rangle \langle \alpha | \beta \rangle=2\Big\langle (\alpha^\dagger \beta)^\dagger (\alpha^\dagger \beta) \Big\rangle_{0+3}=2\Big\langle u_+ \hat \bm_b \hat \bm_a u_+ \hat \bm_a \hat \bm_b u_+ \Big \rangle_{0+3}  \] 
\beq =2\langle \hat \bm_b \hat \bb_+ \hat \ba_+ \hat \bb_+ \hat \bm_b \rangle_{0+3}
=\langle(1+\hat \ba \cdot \hat \bb)u_+\rangle_{0+3}=
\frac{1}{2}(1+\hat \ba \cdot \hat \bb). \label{probadotb} \eeq

This relationship can be more directly expressed in terms of $\bm_a $ and $\bm_b$. 
Using (\ref{ahateqn2}), for $\hat \ba = \frac{2}{\bm_a}-\be_3$ and $\hat \bb=\frac{2}{\bm_b}-\be_3$, a short calculation
gives the result
\beq 1- \frac{(\bm_a-\bm_b)^2}{\bm_a^2 \bm_b^2} = \frac{1}{2}(1+\hat \ba \cdot \hat \bb)
\quad \iff \quad   \frac{(\bm_a-\bm_b)^2}{\bm_a^2 \bm_b^2} = \frac{1}{2}(1-\hat \ba \cdot \hat \bb), \label{probadotb2} \eeq
showing that the probability of finding the particle in that Pauli g-state $|\beta \rangle$
is directly related to the Euclidean distance between the points $\bm_a$ and $\bm_b$.

Clearly, when $\hat \bb = -\hat \ba$, the expression in (\ref{probadotb2}) simplifies to
\[  \frac{(\bm_a-\bm_b)^2}{\bm_a^2 \bm_b^2} = 1. \]
This will occur when $\bm_b := \frac{1}{\bm_a \w \be_3}\bm_a$, for which case
\[  \hat \bb = \hat \bm_b \be_3 \hat \bm_b =
-\hat \bm_a \be_3 \hat \bm_a=-\hat \ba \quad {\rm and} \quad \bm_b \cdot \bm_a = 0. \] 
Writing $\bm_a=\bx_a+\be_3$ for $\bx_a\in \R^2$,
\beq \bm_b=\bx_b+\be_3=\frac{1}{\bx_a \be_3}(\bx_a+\be_3)
=\frac{\be_3 \bx}{\bx^2}(\bx_a+\be_3)=-\frac{1}{\bx_a}+\be_3 .\label{perpma} \eeq
More details of the constructions found in this section can be found in \cite{S2014} and \cite{S1/2}.

\section{Dirac spinors}

What is missing in the concept of a Pauli spinor is the ability to distinguish between Pauli spinors in different
reference frames. Within the geometric algebra $\G_3$, we are able to distinguish different inertial systems.
Recall that the rest frame that defined the geometric algebra $\G_3$ was $\{\be_1, \be_2, \be_3\}$, oriented
by the property that $\be_{123}=\be_1 \be_2 \be_3 = I$. A set of three orthonormal geometric numbers 
$\{\be_1^\prime, \be_2^\prime, \be_3^\prime\}$, specified by the condition 
\[\{\be_1^\prime, \be_2^\prime, \be_3^\prime\} =e^{\frac{1}{2} \phi \hat \bv}\{\be_1, \be_2, \be_3\} e^{-\frac{1}{2} \phi \hat \bv}, \]
where each $\be_k^\prime:=e^{\frac{1}{2} \phi \hat \bv}\be_k e^{-\frac{1}{2} \phi \hat \bv}$, defines
a {\it rest frame}, or {\it inertial system} of {\it relative vectors} moving with a velocity of $\frac{\bv}{c}=\hat \bv\tanh \phi$,
with respect to the inertial system defined by $\{ \be_k\}_{k=1}^3$. If $X=ct+\bx$ represents the {\it time} $t$ and {\it  position vector} $\bx$ of an {\it event} in the inertial system $\{ \be_k\}_{k=1}^3$, then the corresponding event $X^\prime=ct^\prime+\bx^\prime$ in the inertial system $\{ \be_k^\prime\}_{k=1}^3$ is specified by the
{\it active Lorentz transformation} $X^\prime = X e^{-\phi \hat \bv}$, \cite{SNF}, \cite{S81}.

 Each inertial system is distinguished by how
its observer partitions its geometric numbers into vectors and bivectors, in the same sense that what one observer identifies as a pure electric field, becomes a mixture of an (vector) electric and (bivector) magnetic field in an inertial system not at rest. But all
observers have the same pseudoscalar element $I=\be_1 \be_2 \be_3 = \be_1^\prime \be_2^\prime \be_3^\prime$,
representing the unit trivector of space. The {\it relative geometric algebra} 
\[ \G_3^\prime :=gen_{\R} \{\be_k^\prime\}_{k=1}^3=e^{\frac{1}{2} \phi \hat \bv}\G_3 e^{-\frac{1}{2} \phi \hat \bv} ,\]
consists of the same geometric numbers as in $\G_3$, but with a different {\it observer dependent partition} of what elements are identified as (relative) vectors, and what elements are identified as (relative) bivectors. These ideas have been explored by the
author more fully in \cite{S81}, and in the book \cite[Chps:2,11]{SNF}.

Consider now {\it Dirac spinors} of the form
\beq \bM_{\phi}:=e^{\phi \be_3}\bx+\be_3 =\bx \cosh \phi + \be_3 + I \be_3 \times \bx \sinh \phi \label{diracspinorform} \eeq
where $\bx=x_1 \be_1+x_2 \be_2$, and  $\phi \in \R$.
When $\phi = 0$, $\bM_{0}=\bx + \be_3$ becomes the Pauli spinor given in (\ref{mnbx}).  Indeed, since $\bM_{\phi}^2=\bx^2+1\ge 1$,
\beq \hat \bM_\phi =\frac{\bM_\phi}{\sqrt{\bx^2+1}} =  e^{\frac{1}{2} \phi \be_3}\frac{\bM_0}{\sqrt{\bx^2+1}}e^{-\frac{1}{2} \phi \be_3},  \label{Mzeroform} \eeq
so $\hat\bM_\phi$ is just the normalized spinor defined by $\alpha = \hat \bm u_+$ boosted into the inertial system $ \{\be_k^\prime\}_{k=1}^3$ 
where $\hat \bv=\be_3$.

Starting with (\ref{diracspinorform}), we calculate
\beq \hat \bM_{\phi}=  \frac{\bx \cosh \phi+\be_3}{\sqrt{\bx^2+1}}+ 
  \frac{I(\be_3\times \bx)\sinh \phi}{\sqrt{\bx^2+1}} = \bm_1+I \bm_2
   = e^{\frac{1}{2}\omega \hat \bm_1\times \hat \bm_2} \hat \bm_1 e^{-\frac{1}{2}\omega \hat \bm_1\times \hat \bm_2}   , \label{m1m2form} \eeq
where $\cosh \omega :=|\bm_1|$ and $\sinh \omega := |\bm_2|$.
which defines the velocity 
\[ \tanh\frac{\omega}{2} ={\frac{v_\omega}{c}}:= \sqrt{\frac{(\bm_2)^2}{(\bm_1)^2}}= \frac{|\bx| \sinh \phi}{\sqrt{\bx^2 \cosh^2 \phi+1}}  \]
in the direction
\[  \hat \bm_1 \times \hat \bm_2 = \frac{-\hat\bx + \be_3|\bx|\cosh \phi }
{\sqrt{\bx^2 \cosh^2 \phi+1}} . \]
From (\ref{Mzeroform}) and (\ref{m1m2form}), it follows that the spinor $\bM_0$ boosted in the direction of $\be_3$, with
rapidity $\tanh \phi = \frac{v}{c}$, is in the same inertial system as the spinor defined by $\hat \bm_1$ boosted in the
direction of $\hat \bm_1\times \hat \bm_2$ with rapidity $\tanh \omega = \frac{\omega}{c}$.

Just as the equations (\ref{ahateqn}) and (\ref{ahateqn2}) led to the interpretation of stereographic projection on the Riemann sphere, Figure \ref{sterox}, the generalized
equation
\[  \bM_\phi = \frac{2}{\hat \bA+\be_3}\quad \iff \quad \hat \bA = \hat \bM_\phi \be_3 \hat \bM_\phi, \]
where $\hat \bA^2=1$,
 leads to a stereographic projection of the Riemann sphere, followed by a boost. This is most clearly seen by
writing 
\[ \hat \bM_\phi \be_3 \hat \bM_\phi = e^{\frac{1}{2}\phi \hat \bm_1\times \hat \bm_2}\hat \bm_1\be_3 \hat \bm_1 
 e^{-\frac{1}{2}\phi \hat \bm_1\times \hat \bm_2},\]
so the Dirac $g$-spinor represented by $\hat \bM_\phi$, maps the North Pole $\be_3$ into the point 
$\hat \ba = \hat \bm_1 \be_3 \hat \bm_1$ on the Riemann sphere, followed by the boost $e^{\frac{1}{2}\phi \hat \bm_1\times \hat \bm_2}$. See \cite{S2015} for more details of this construction.


\begin{thebibliography}{}
\bibitem{BK12} W.E. Baylis, J.D. Keselica, {\em The Complex Algebra of Physical Space: A Framework for Relativity},
{\it Advances in Applied Clifford Algebras}, Vol. 22, No. 3, pp. 537 - 561, 2012.
\bibitem{H99} D. Hestenes, {\em New Foundations for Classical Mechanics, 2nd Ed.}, Kluwer 1999.
\bibitem{H10} D. Hestenes, {\em Zitterbewegung in Quantum Mechanics}, Found Physics (2010) 40:1-54.
\begin{verbatim} http://geocalc.clas.asu.edu/pdf/ZBWinQM15**.pdf \end{verbatim}
\bibitem{S2014} G. Sobczyk, {\em Part I: Vector Analysis of Spinors}, (2014) 
\begin{verbatim} http://arxiv.org/abs/1507.06608 \end{verbatim}
\bibitem{S2015} G. Sobczyk, {\em Part II: Spacetime Algebra of Dirac Spinors}, (2015)
\begin{verbatim} http://arxiv.org/abs/1507.06609 \end{verbatim}
\bibitem{Sob2012} G. Sobczyk, {\it Conformal Mappings in Geometric Algebra}, Notices of the AMS, 
Volume 59, Number 2, p.264-273, 2012.
 \bibitem{S1/2} G. Sobczyk, {\em Geometry of Spin 1/2 Particles}, Revista Mexicana de F\'isica, {\bf 61} (2015) 211-223.
\begin{verbatim} http://rmf.smf.mx/pdf/rmf/61/3/61_3_211.pdf \end{verbatim}
\bibitem{SNF} G. Sobczyk, {\em New Foundations in Mathematics: The Geometric Concept of Number},
\newblock Birkh\"auser, New York 2013.
\bibitem{S08} G. Sobczyk, Geometric Matrix Algebra, {\em Linear Algebra and its Applications}, 429 (2008) 1163-1173.
\bibitem{S1} G. Sobczyk, Hyperbolic Number Plane, {\it The College Mathematics
Journal}, Vol. 26, No. 4, pp.268-280, September 1995.
 \bibitem{S81} G. Sobczyk, {\em Spacetime Vector Analysis}, Physics Letters A, Vol 84A, p.45-49, 1981.
\end{thebibliography}
\end{document}